\documentclass[superscriptaddress,aps,prd,twocolumn,showpacs,floatfix,preprintnumbers,amsmath,amssymb,nofootinbib,groupedaddress]{revtex4}

\input epsf
\usepackage{graphicx}
\usepackage{color}
\usepackage{mathtools}

\def\VEV#1{\left\langle #1 \right\rangle}

\newcommand{\om}{\omega}

\newcommand{\be}{\begin{equation}}
\newcommand{\ee}{\end{equation}}
\newcommand{\ba}{\begin{align}}
\newcommand{\ea}{\end{align}}

\newcommand{\fsky}{f_{\rm sky}}

\newcommand{\Om}{\Omega}

\begin{document}
			
			\title{Search for Compensated Isocurvature Perturbations with Planck Power Spectra}
			
			\author{Julian B. Mu\~noz}  \email{julianmunoz@jhu.edu}
			\affiliation{Department of Physics and Astronomy, Johns
				Hopkins University, 3400 N.\ Charles St., Baltimore, MD 21218}
			\author{Daniel Grin}  \email{dgrin@kicp.uchicago.edu }
			\affiliation{Kavli Institute for Cosmological Physics, University of Chicago, Chicago, IL 60637}
			\affiliation{Department of Astronomy and Astrophysics, University of Chicago, Chicago, IL 60637}
			\author{Liang Dai}  
			\affiliation{Institute for Advanced Study, Einstein Drive, Princeton, New Jersey 08540}
			\author{Marc Kamionkowski}  
			\affiliation{Department of Physics and Astronomy, Johns
				Hopkins University, 3400 N.\ Charles St., Baltimore, MD 21218}
			
			\author{Ely D. Kovetz}  
			\affiliation{Department of Physics and Astronomy, Johns
				Hopkins University, 3400 N.\ Charles St., Baltimore, MD 21218}
			
			\date{\today}
			
			\begin{abstract}
				In the standard inflationary scenario, primordial perturbations are adiabatic.  The amplitudes of most types of isocurvature perturbations are generally constrained by current data to be small.  If, however, there is a baryon-density perturbation that is compensated by a dark-matter perturbation in such a way that the total matter density is unperturbed, then this compensated isocurvature perturbation (CIP) has no observable consequence in the cosmic microwave background (CMB) at linear order in the CIP amplitude.  Here we search for the effects of CIPs on CMB power spectra to quadratic order in the CIP amplitude. An analysis of the Planck temperature data leads to an upper bound $\Delta_{\rm rms}^2 \leq 7.1\times 10^{-3}$, at the 68\% confidence level, to the variance $\Delta_{\rm rms}^2$ of the CIP amplitude.  This is then strengthened to $\Delta_{\rm rms}^2\leq 5.0\times 10^{-3}$ if Planck small-angle polarization data are included.  A cosmic-variance-limited CMB experiment could improve the $1\sigma$ sensitivity to CIPs to $\Delta^2_{\rm rms} \lesssim 9\times 10^{-4}$. It is also found that adding CIPs to the standard $\Lambda$CDM model can improve the fit of the observed smoothing of CMB acoustic peaks just as much as adding a non-standard lensing amplitude.			
			\end{abstract}
			\maketitle
			
			\section{Introduction}
			
			Cosmological density perturbations are thought to have their origin during inflation \cite{Mukhanov:1985rz,astro-ph/0504097,Bardeen:1983qw}. These perturbations seed the large-scale inhomogeneities that later grow to become galaxies and clusters \cite{astro-ph/9506072}. From structure formation and cosmic-microwave-background (CMB) 
			observations, we know that primordial density perturbations have amplitude $\delta \sim 10^{-5}$ \cite{1502.01589}. 
			
			Primordial perturbations can be classified into two groups depending on their initial conditions. Adiabatic perturbations are perturbations to the total energy density that leave the ratios of the different constituents of matter everywhere the same.  Isocurvature perturbations involve perturbations to the relative number densities of different components of matter \cite{Peebles:1970ag,astro-ph/9610219,astro-ph/0009131,astro-ph/9809142}. The simplest inflationary models have purely adiabatic fluctuations, while isocurvature fluctuations usually signal the presence of a second field during inflation, as in curvaton models \cite{hep-ph/0110002,astro-ph/0208055}.
			
			Isocurvature perturbations between photons and a single other species
			are in general well constrained \cite{astro-ph/0409326,astro-ph/0509209,0705.2853}. 
			If, however, there is a baryon-density perturbation that is compensated by a dark-matter perturbation in such a way that the total matter density remains constant, then there are no pressure or gravitational-potential perturbations above the baryonic Jeans scale.  These compensated isocurvature perturbations (CIPs) thus have no observable effect on the CMB at linear order in the CIP amplitude \cite{astro-ph/0212248,0907.5400}.  There are constraints from other observables, but these are a factor of $\sim 10^4$ weaker than the adiabatic component \cite{0907.3919,1107.5047,1107.1716,1306.4319}. 
			
			In the standard scenario, the CMB power spectrum is determined given fixed values of the baryon and dark-matter densities $\Omega_b$ and $\Omega_m$, respectively, in units of the critical density.  CIPs, however, introduce variations to $\Omega_m$ and $\Omega_b$ between different patches of the CMB sky.  They thus induce a variation in the power spectrum from one patch of sky to another.  
			
			The {\it mean} power spectrum---that obtained by measurements over the entire sky---remains unaltered, to linear order in the CIP amplitude.  The variations show up, however, in two different ways.  First of all, the spatial modulation of the power spectrum is characterized by a departure from gaussianity, a specific nontrivial four-point function, or trispectrum.  A search for such a trispectrum was performed in Ref.~\cite{1306.4319}.  The second consequence, however, is a change to the power spectrum that arises to quadratic order in the CIP amplitude, which can be understood heuristically as a smoothing of features in the CMB power spectrum when different power spectra are averaged.
			
			In this paper we seek this effect of CIPs on the CMB power spectra obtained by Planck. We parametrize the magnitude of the effect of CIPs in terms of an rms CIP amplitude $\Delta_{\rm rms}$.  We find from a temperature-only analysis a constraint of $\Delta_{\rm rms}^2\leq 7.1\times 10^{-3}$, which is competitive with, and complements, that obtained from the complete trispectrum, although with a far simpler analysis. That figure improves to $\Delta_{\rm rms}^2\leq 5.0\times 10^{-3}$ if Planck polarization data are included.
			We then make CIP sensitivity forecasts for future experiments.  We also show that CIPs have a very similar effect on the power spectrum to changing the lensing amplitude $A_L$. They can thus
			alleviate the tension between the lensing amplitude obtained from the Planck spectrum ($A_L=1.22\pm0.1$) and that expected from theory ($A_L=1$) \cite{1502.01589}. 
			
			This paper is structured as follows. In Section~\ref{sec:cips} we review the physics of CIPs and their effects on the CMB power spectra. Then, in Section~\ref{sec:estimators} we find a linear basis of estimators for CMB observables, including CIPs. In Section~\ref{sec:CMB} we apply our analysis to current CMB data and an ideal cosmic-variance-limited CMB experiment. We conclude in Section ~\ref{sec:conclusions}.
			
			\section{Compensated Isocurvature Perturbations}
			\label{sec:cips}
			
			CIPs change the baryon and dark-matter densities in such a way that the total matter energy density, $\Omega_m=\Omega_b+\Omega_c$, remains unaltered. We  parametrize their effect as
			\ba
			\Omega_b &=\bar \Omega_b [1+\Delta (\hat n)], \quad \rm{and} 
			\nonumber \\ 
			\Omega_c &= \bar \Omega_c -  \bar \Omega_b \Delta(\hat n),
			\label{eq:Delta}
		\end{align}
		where $\Omega_b$ and $\Omega_c$ are the baryon and dark-matter energy densities respectively, the overbar represents their unperturbed values, and $\Delta (\hat n)$ is the amplitude of the CIP in the specific direction $\hat n$ at recombination. This expression is accurate for CIPs of sufficiently large angular scale, where they can be treated as a modulation of background parameters \cite{1505.00639}.

		The linear-order effects of CIPs are on scales at which the baryons behave differently from dark matter, corresponding to angular scales $\ell\gtrsim 10^{5-6}$ \cite{1107.5047}, which makes them unobservable in the CMB, although potentially detectable using cosmological 21-cm absorption measurements at high redshift \cite{0907.5400}.
		CIPs will also have consequences for the CMB fluctuations induced by adiabatic perturbations.  
		In a region of high $\Delta(\hat n)$ (high baryon density), decoupling will be longer, thereby smoothing the peak structure in the CMB power spectrum. The mean-free path of CMB photons would be reduced by the higher electron density, leading to less damping of CMB fluctuations on small angular scales.
		A higher baryon density also decreases the plasma sound speed and hence decreases the sound horizon at recombination \cite{Peebles:1970ag}.
		
		\subsection{Angular properties}
		
		We expand the amplitude of the compensated isocurvature perturbations in spherical harmonics as
		\be
		\Delta(\hat n) = \sum_{LM} Y_L^M(\hat n) \Delta_{LM},
		\ee
		where the spherical-harmonic coefficients $\Delta_{LM}$ are statistically independent and have a variance given by
		\be
		\VEV{\Delta_{LM}\Delta_{L'M'}^{*}} = \delta_{LL'} \delta_{MM'}  C_L^\Delta.
		\ee
		We take the \emph{ansatz} of a scale-invariant power spectrum for $\Delta$ in $k$-space. For $L\lesssim 800$, this creates a scale-invariant angular power spectrum $C_L = A \, L^{-2}$ when projected onto the last-scattering surface (LSS), where $A$ is a dimensionless amplitude \cite{1107.5047}. 
		The simple picture of CIPs as a modulation of background parameters corresponds 
		to a separate universe approximation, which was shown in Ref.~\cite{1505.00639} to only be valid for $L\lesssim100$, as the imprint of CIPs are washed for at smaller CIP angular scales.
		We thus restrict our analysis to $L \leq 100$.
		
		We assume that the CIP amplitude $\Delta(\hat n)$ is a Gaussian random variable with zero average and variance $\Delta^2_{\rm rms}\equiv \VEV{\Delta^2}$.  Instead of finding an estimator for each $\Delta_{LM}$ we will directly measure its variance, which can be expressed in terms of the CIP angular power spectrum $C_L^\Delta$ as
		\be
		\Delta^2_{\rm rms} = \sum_{L=1}^{100} \dfrac{(2L+1)}{4\pi}  C_L^\Delta,
		\label{eq:ClDelta}
		\ee
		which means  that our constraints will be on the total power in CIPs and not on each individual $C_L^\Delta$. By using Eq.~(\ref{eq:ClDelta}) we can relate the CIP variance $\Delta^2_{\rm rms}$ to the amplitude $A$ of the power spectrum as,
		\be
		\Delta^2_{\rm rms} \approx 0.96 A.
		\ee
		
		\subsection{Previous constraints}
		
		As CIPs do not change CMB power spectra at linear order, past work has relied on other observables to constrain their amplitude. 
		For example, measurements of galaxy-cluster baryon fractions (obtained through X-ray observations) were used in Ref.~\cite{0907.3919} to search for CIPs, imposing the constraint $\Delta_{\rm rms}^{2}\lesssim 5\times 10^{-3}$.
		This technique, however, relies on clusters being fair samples of the baryon density in the universe, as well as being kinematically relaxed.
		
		In Ref.~\cite{1107.5047}, the off-diagonal correlations in the CMB created by CIPs were computed, and a forecast was made of the sensitivities that could be reached by studying them with different instrumental setups. 
		Data from the WMAP mission \cite{1212.5226} were analyzed in Ref.~\cite{1306.4319} to constrain the amplitude $A$ of the CIP power spectrum $C_L$ to be smaller than $5.5\times 10^{-3}$ at 68\% C.L., which translates to a constraint on the CIP variance of $\Delta^2_{\rm rms} \lesssim 4 \times 10^{-3}$, where the $L=1$ mode has been ignored due to reconstruction uncertainties.
		
		
		\subsection{Effect on the power spectrum}
		\label{sec:newcont}

		In our picture we treat the CIP amplitude as a Gaussian random variable. This allows us to calculate the observed CMB angular power spectrum $C_\ell^{\rm obs}$ by averaging over the CIP amplitudes,
		\be
		C_\ell^{\rm obs} = \dfrac 1 {\sqrt{2\pi\Delta^2_{\rm rms}}} \int d \Delta e^{-\Delta^2/(2\Delta_{\rm rms}^2)} C_\ell(\Delta),
		\ee
		which to first non-zero order in $\Delta_{\rm rms}$ is given by
		\be
		C_\ell^{\rm obs} \approx C_\ell|_{\Delta = 0} + \dfrac 1 2 \dfrac{d^2 C_\ell}{d \Delta^2} \Delta^2_{\rm rms}.
		\label{eq:dCL}
		\ee
		
		We calculate the second derivative by fitting $C_\ell|_{\Om_b,\Om_c}$ near $\{\bar \Omega_b$, $\bar \Omega_c\}$
		as a function of $\Delta$. We have checked that terms that are higher order in $\Delta_{\rm rms}^2$ are negligible for the upper limits to $\Delta_{\rm rms}^2$ that we infer.
		
		We have found that the CIP-induced corrections to CMB power spectra from Eq.~(\ref{eq:dCL}) numerically agree with those computed with the full mode-coupling formalism of Ref.~\cite{1107.5047}, but are simpler to evaluate.

		\section{CIP Estimators}
		\label{sec:estimators}
		
		We have shown expressions for how CIPs alter CMB power spectra. Now we consider how to estimate $\Delta^2_{\rm rms}$ with measurements of the three main CMB power spectra, $C_\ell^{TT}$, $C_\ell^{TE}$, and $C_\ell^{EE}$. For that we use a Fisher-matrix analysis to fully capture the correlations between the CIP variance, $\Delta^2_{\rm rms}$, and the usual cosmological parameters.
		
		Let us begin by reviewing the basics of linear (Fisher) cosmological-parameter estimation.

		\subsection{Linear estimators}
		
		Codes like CosmoMC \cite{astro-ph/0205436} and Python Monte Carlo \cite{1210.7183} are commonly used for parameter analysis. It is, however, a computationally costly procedure.
		We already have a best fit for the six $\Lambda$CDM model parameters in the absence of CIPs \cite{1502.01589}, so we can perturb the model around this best fit by adding CIPs, increasing the number $N_p$ of parameters to seven. In that case the new best-fit parameters will not be too far away in parameter space from the old ones, so we can perform a linear analysis.

		We construct a linear estimator of the parameters near their current best-fit values \cite{1502.01589}. To do so we parametrize the power spectra as,
		\be
		C_\ell^{X,\rm obs} - C_\ell^{X,\rm best-fit} =\sum_{i=1}^{N_p} \delta A^{X}_i g^{X}_i(\ell), 
		\ee
		where $C_\ell^{X,\rm best-fit}$ is the best-fit (lensed) power spectrum, with $X=\{TT, TE, EE\}$. 
		We have left out other CMB observables, such as B-mode polarization, due to the absence of sufficiently sensitive and foreground-free CMB polarization data. These could potentially have significant constraining power \cite{0907.3919,1107.5047}.
		
		We define the first six original amplitudes to be the $\Lambda$CDM parameters as $A_i = \{\omega_b,\omega_c,n_s-1,A_s,\tau,H_0\}$, where $\omega_b=\Omega_b h^2$ and $\omega_c=\Omega_c h^2$ are the baryon and cold-dark-matter physical densities, $n_s$ is the tilt of the scalar power spectrum, and $A_s$ its amplitude. Here, $\tau$  is the optical depth of reionization and $H_0$ is the Hubble parameter. 
		We define the deviations of these parameters from their best-fit values to be $\delta A_i$.
		
		The basis functions $g^{X}_i(\ell)$ for $i=1-$6 are constructed as
		\be
		g^{X}_i(\ell) \equiv \dfrac{\partial C^{X}_\ell}{\partial A_i},
		\label{eq:gi}
		\ee
		where the derivatives are taken by fitting in CAMB \cite{astro-ph/9911177} near the best-fit values of the six $\Lambda$CDM parameters.
		
		Since we are going to add CIPs we will have $N_p=7$, unless otherwise specified.
		The change in the power spectrum when adding CIPs is parametrized by Eq.~(\ref{eq:dCL}), from where we can extract the seventh basis function
		\be
		g_7^{X}(\ell) \equiv  \dfrac 1 2 \dfrac{d^2 C^{X}_\ell}{d \Delta^2},
		\ee
		where the derivative is calculated in the separate-universe approximation, and has associated amplitude $\delta A^{X}_7 = \Delta_{\rm rms}^2$. 
		
\begin{figure}
			\includegraphics[width=88mm]{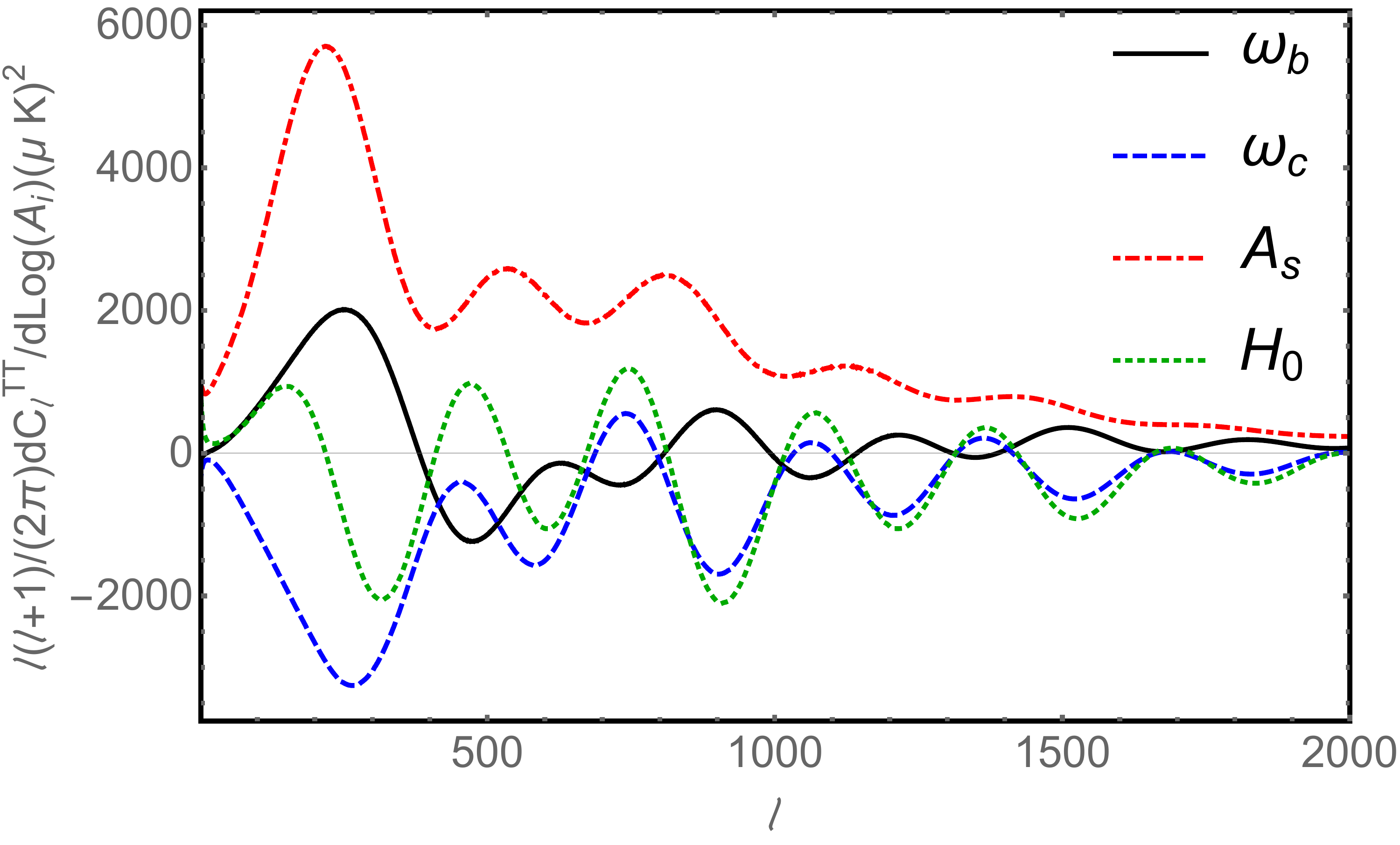}
			\\
			\includegraphics[width=88mm]{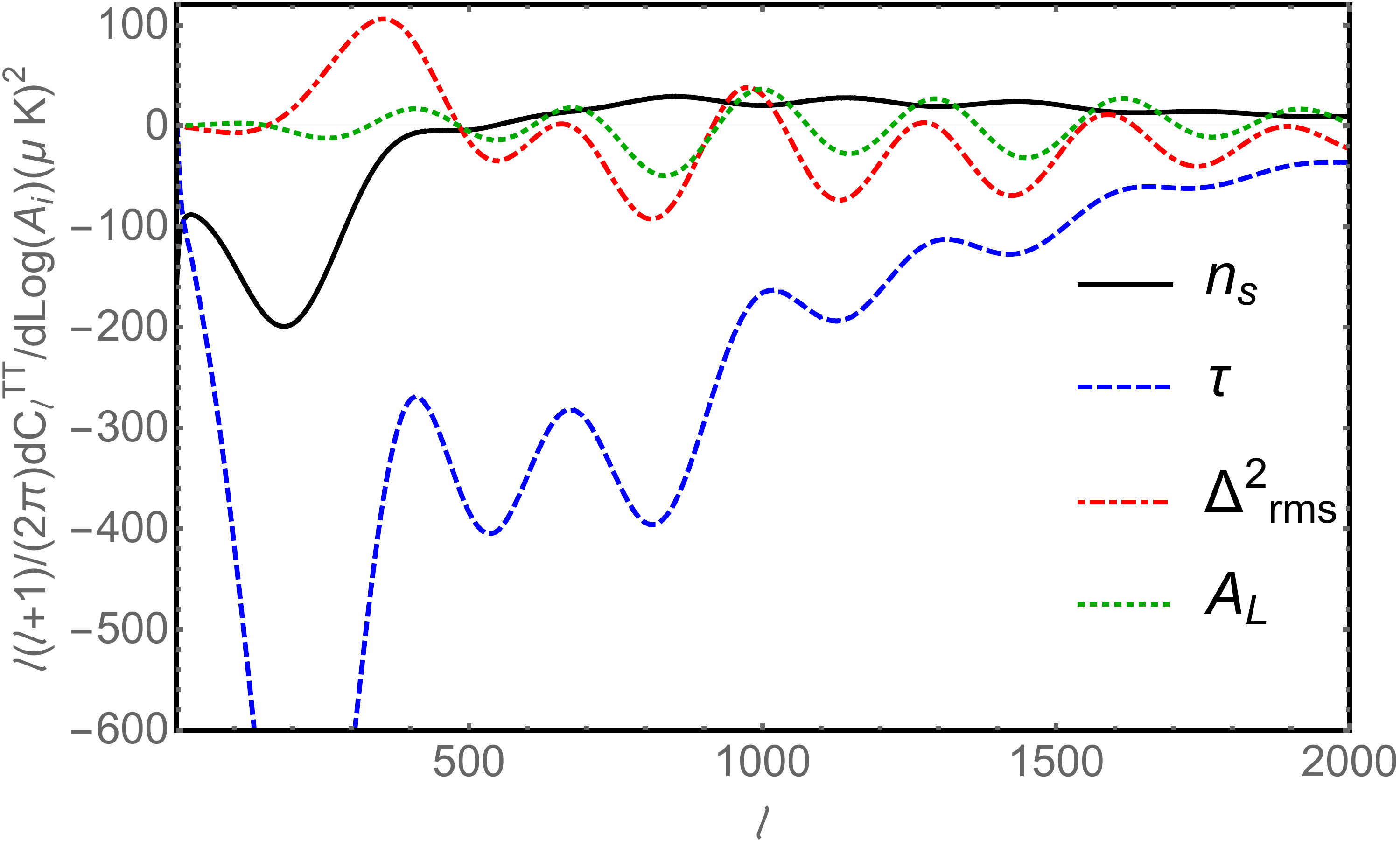}
			\caption{Derivatives of the CMB TT power spectrum at the current best-fit values. We employ derivatives with respect to the logarithm of each amplitude $A_i$ to account for their different orders of magnitude. Consequently, there is a factor of $A_i$ different to translate to the $g_i$s in the text. In the top panel we show the derivatives with respect to $\omega_b$ (in solid-black), $\omega_c$ (in dashed-blue), $A_s$ (in red--dot-dashed), and $H_0$ (in dotted-green). In the lower panel we plot the derivatives with respect to $n_s$ (in solid-black), $\tau$ (in dashed-blue), the CIP variance $\Delta^2_{\rm rms}$ (in red--dot-dashed), and the lensing amplitude $A_L$ (in dotted-green). For visual purposes we have chosen an arbitrary CIP normalization in these plots.}
			\label{fig:gi}
		\end{figure}

\begin{figure}
			\includegraphics[width=88mm]{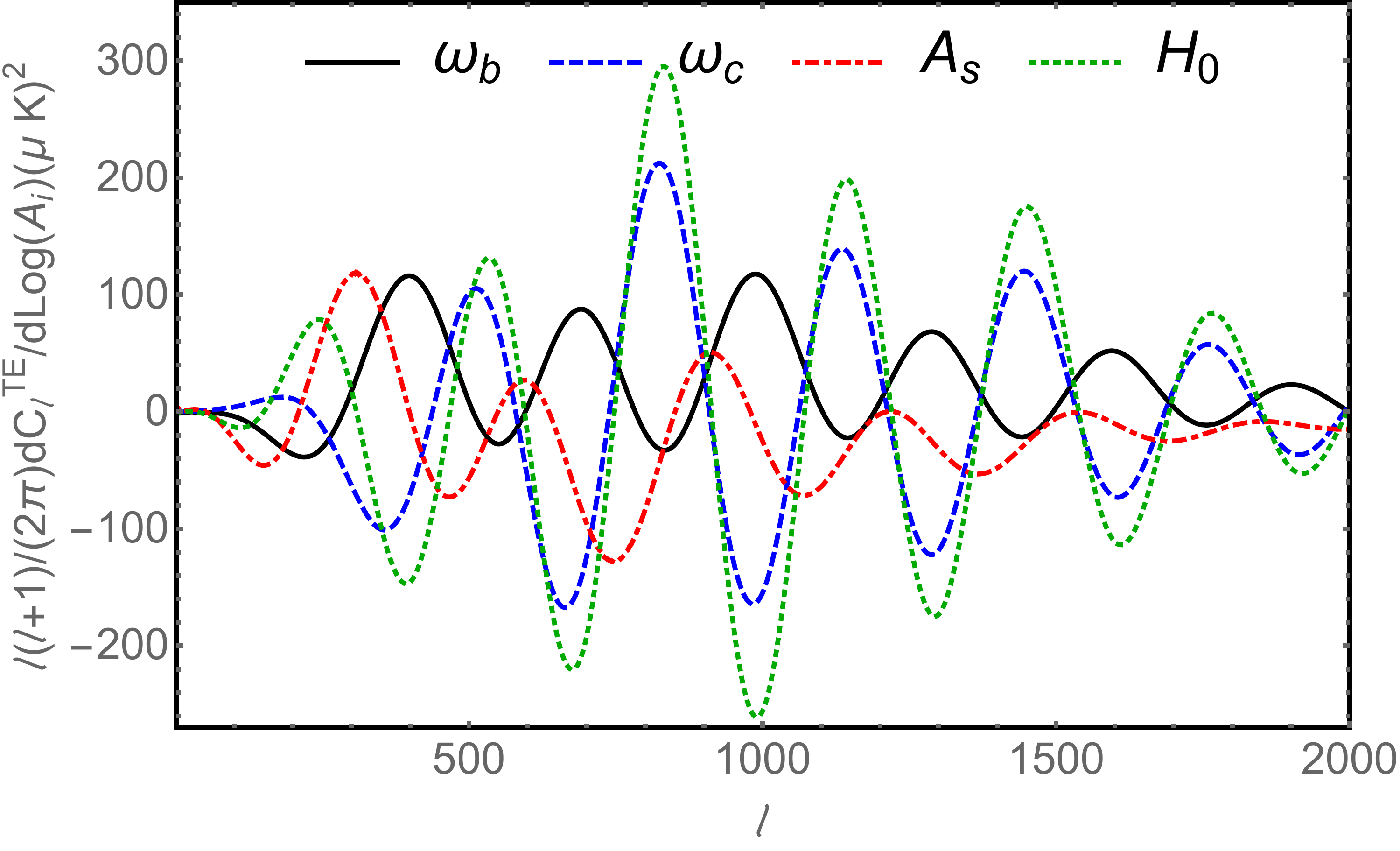}
			\\
			\includegraphics[width=88mm]{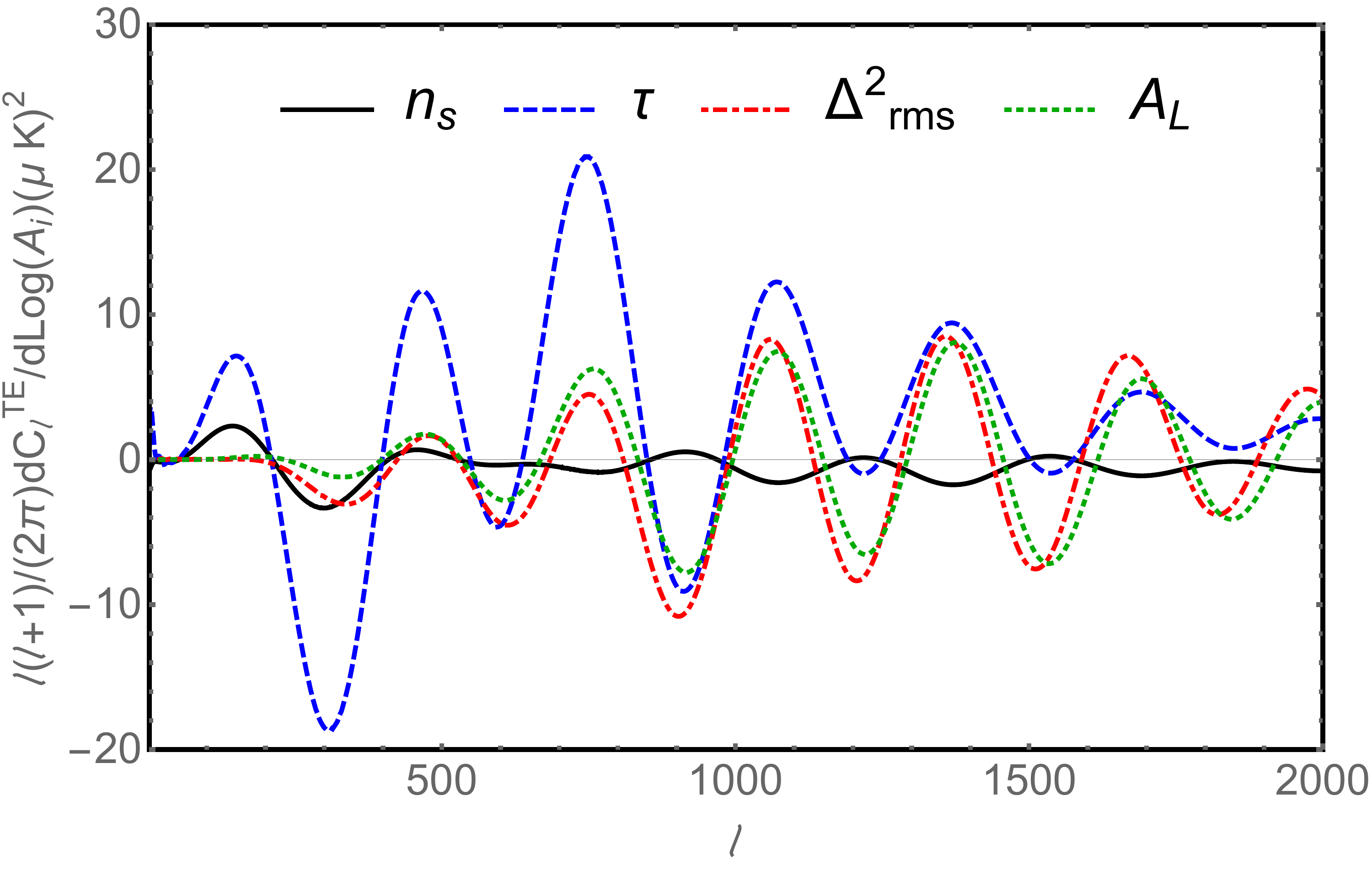}
			\caption{Derivatives of the CMB TE power spectrum at the current best-fit values. We use the same conventions as in Figure~\ref{fig:gi}.}
			\label{fig:giTE}
		\end{figure}

\begin{figure}
			\includegraphics[width=88mm]{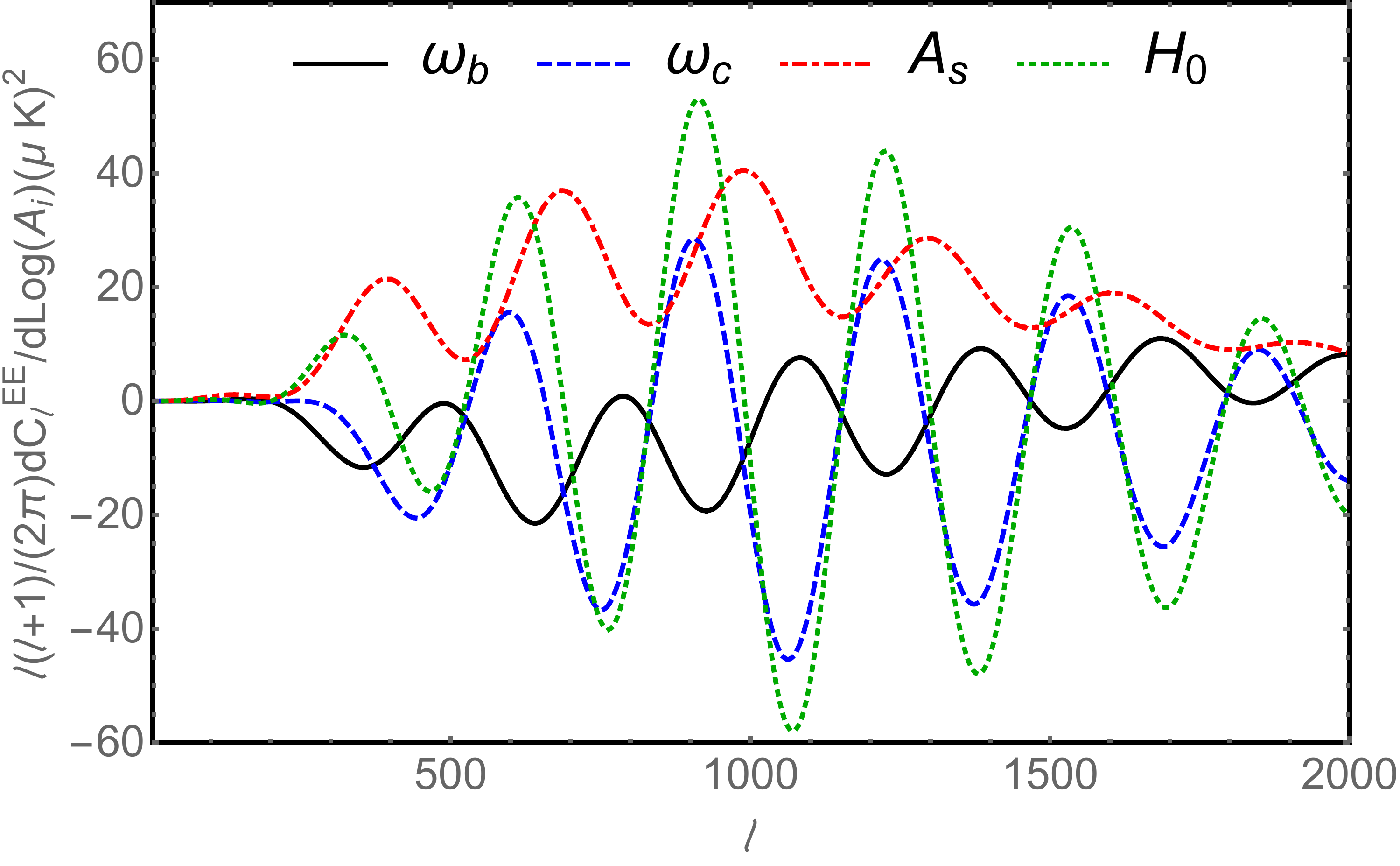}
			\\
			\includegraphics[width=88mm]{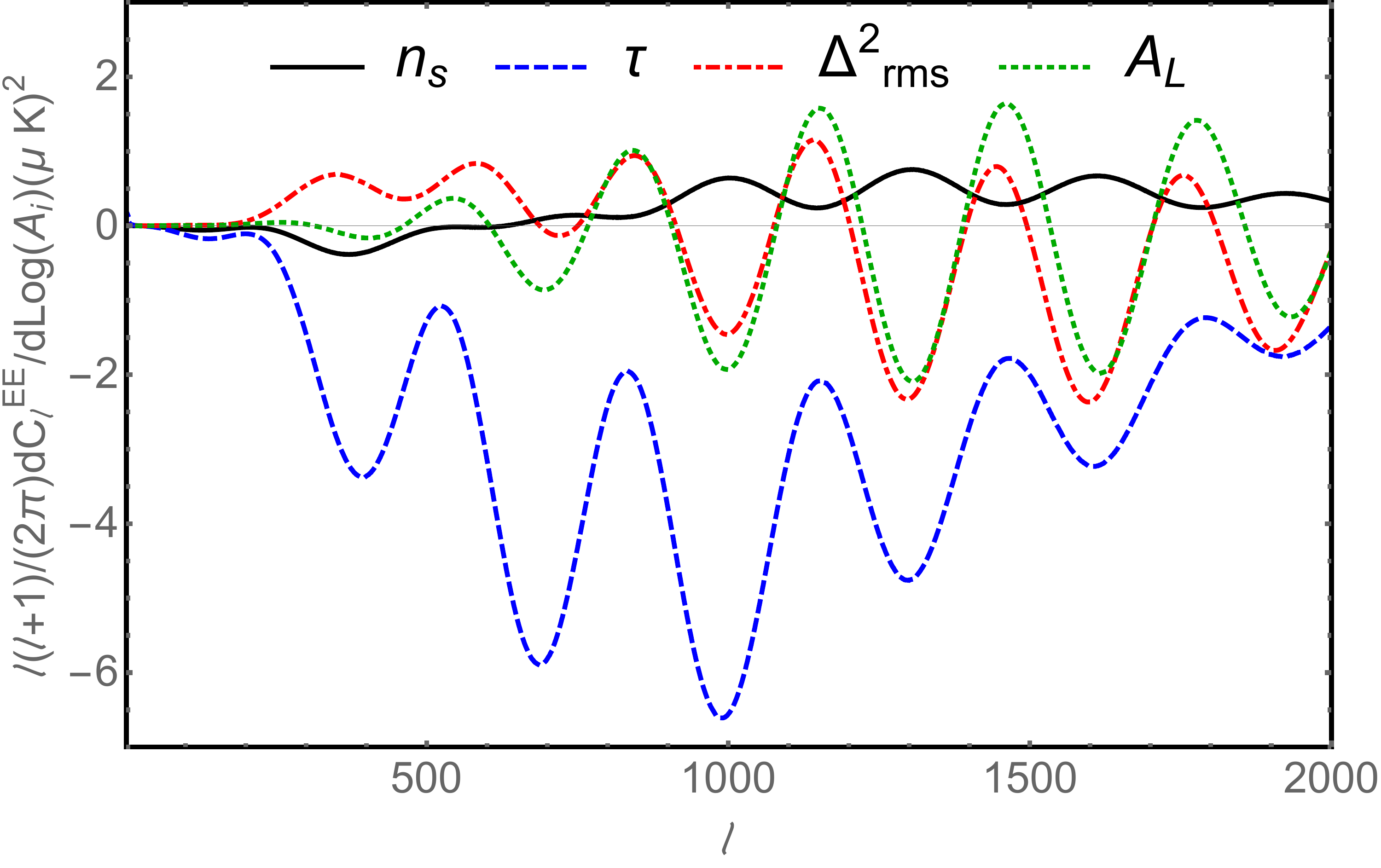}
			\caption{Derivatives of the CMB EE power spectrum at the current best-fit values. We use the same conventions as in Figure~\ref{fig:gi}.}
			\label{fig:giEE}
		\end{figure}

		We show all the derivatives in Figures~\ref{fig:gi}, \ref{fig:giTE}, and \ref{fig:giEE}. There are well-known correlations between the high-$\ell$ effects of changing the dark-matter density $\om_c$ and the Hubble parameter $H_0$. Similarly, increasing $A_s$ and decreasing $\tau$ produce very similar changes in the power spectra, except at the lowest $\ell$s. 
		
		Notice that in those plots we are also showing the derivative with respect to the lensing amplitude as an eighth parameter.  The basis functions for CIPs and lensing are very similar. This could help resolve the tension between the observed level of CMB lensing in Planck power spectra and expectations from the $\Lambda$CDM model. We will explore this topic in Section~\ref{sec:CMB}.
		
		\subsection{Fisher Matrix}
		We now study the detectability of the seven $\delta A_i$ simultaneously through a Fisher analysis. We employ the usual definition of the Fisher matrix \cite{Jungman:1995bz,astro-ph/9603021,1403.5271}, with components
		\be
		F_{ij} = \VEV{g_i,g_j},
		\ee
		where the inner product $\VEV{\,,}$ is defined as
		\be
		\VEV{g_i,g_j} \equiv  \sum_{X,Y} \sum_\ell g^X_i(\ell)\mathcal C^{-1}_{XY}g^Y_j(\ell).
		\ee
		The covariance matrix $\mathcal C_\ell$ is given by \cite{astro-ph/9611125,astro-ph/9609170}
		\begin{equation}
		(\mathcal C_\ell)_{XY} = \dfrac{2}{2\ell+1} \dfrac{1}{\fsky} \times$$ $$ \times \left ( 
		\begin{tabular}{c c c}
		$\left(\tilde C_\ell^{TT}\right)^2$ \quad & $\left(\tilde C_\ell^{TE}\right)^2$ & $\tilde C_\ell^{TT} \tilde C_\ell^{TE}$ \\
		
		$\left(\tilde C_\ell^{TE}\right)^2$ \quad & $\left(\tilde C_\ell^{EE}\right)^2$ & $\tilde C_\ell^{EE} \tilde C_\ell^{TE}$ \\

		$\tilde C_\ell^{TT} \tilde C_\ell^{TE}$ \quad & $\tilde C_\ell^{EE} \tilde C_\ell^{TE}$ & \small $ \left  [\left(\tilde C_\ell^{TE}\right)^2+\tilde C_\ell^{TT} \tilde C_\ell^{EE}\right] \dfrac 1 2$ \\
		\end{tabular}
		\right),
		\end{equation}
		where we have defined
		\ba
		\tilde C_\ell^{TT} &\equiv C_\ell^{TT} + N_\ell^{TT}, \nonumber \\ 
		\tilde C_\ell^{TE} &\equiv C_\ell^{TE} , \nonumber \\ 
		\tilde C_\ell^{EE} &\equiv C_\ell^{EE} + N_\ell^{EE},
	\end{align}
	and the $N_\ell^X$ are the instrumental noises, for which we use the Planck tabulated noise for the Planck analysis and zero in the cosmic-variance-limited case.

	\section{CMB analysis}
	\label{sec:CMB}
	
	Now we are ready to find estimates and errors for the six standard cosmological parameters, as well as the  CIP amplitude $\Delta_{\rm rms}^2$.

	We consider two cases. First, for Planck, we not only obtain estimators for the CIP variance, but also apply them to the data to obtain actual limits to CIPs. We will take a small detour to study the viability of CIPs as a solution for the lensing tension in the Planck CMB power spectra. 
	Second, we will study a cosmic-variance limited (CVL) experiment.

	\subsection{Planck constraint}
	
	Let us begin by considering the Planck 2015 power spectra ($C_\ell^{X, \rm Planck}$), obtained from the Planck Legacy Archive\footnote{http://pla.esac.esa.int/pla/}. To diminish the effects of correlations between different $\ell$s, we used binned data for $\ell\geq 30$, with width $\Delta \ell=30$.
	The minimum-variance unbiased estimators for these seven amplitudes $\delta A_i$ are
	\be
	\widehat{\delta A}_i  = \sum_j (F^{-1})_{ij} \VEV{R(\ell),g_j(\ell)},
	\label{eq:estimators}
	\ee
	where $(F^{-1})_{ij}$ is the inverse of the Fisher matrix, and $R(\ell)$ is the residual after subtracting the best fit from the data, $R^X(\ell) = C_\ell^{X,\rm Planck} - C_\ell^{X,\rm best-fit}$.

	With the current data in the Planck Legacy Archive, however, it is hard to disentangle the optical depth $\tau$ and the scalar amplitude $A_s$, since the effect of changing either is highly degenerate \cite{Bond:1987ub}.  The main difference between $A_s$ and $\tau$ is the reionization bump, caused by $\tau$, that appears at low $\ell$ in polarization measurements \cite{Ng:1994sv,astro-ph/9608050,astro-ph/0302404}. Our linear analysis underestimates the errors when using low-$\ell$ polarization data, so in lieu of them we will add a prior $\tau=0.068 \pm 0.019$ to the optical depth for robustness. We choose the final $\ell$ ranges to be $\ell=30-2500$ for TT, and $\ell=30-1995$ for TE and EE power spectra, where the maximum $\ell$ is that available in the Planck Legacy Archive. Later on, when considering lensing, we will add the full low-$\ell$ data to the analysis.
	
	We show the best fits derived with this analysis in Table ~\ref{tab:Planck}.
	The best fit to the CIP amplitude with TT-data only is $\Delta^2_{\rm rms}= (5.8 \pm \,7.1) \times 10^{-3}$, and with the combined data set is $\Delta^2_{\rm rms}= (0.9 \pm 5.0) \times 10^{-3}$. There is thus no evidence for the existence of CIPs, and the constraint is of the same order of magnitude as the trispectrum constraint of Ref.~\cite{1306.4319}. Notice that we have not required $\Delta^2_{\rm rms}$ to be positive. Imposing a prior $\Delta^2_{\rm rms}\geq0$ would change the 68\%C.L. constraints to $\Delta^2_{\rm rms}\leq 0.011$ for TT, $\Delta^2_{\rm rms}\leq0.012$ for TE, $\Delta^2_{\rm rms}\leq0.052$ for EE, and $\Delta^2_{\rm rms}\leq0.0054$ for the combined data set. 
	Notice that these limits have become more constringent in the case of the TE data set, due to the negative best-fit value for $\Delta^2_{\rm rms}$, whereas the opposite is true for the TT and EE data sets.
	

	\begin{table*}[hbtp!]
		\begin{tabular}{ l  c  c  c  c  }
			\hline
			\hline
			Parameter &  TT & TE & EE & Combined   \\             
			\hline
			$\omega_b \dots\dots\dots$ & 0.02238$\pm$0.00028 & 0.02175$\pm$0.00052 & 0.0251$\pm$0.0015 & 0.02243$\pm$0.00017 \\
			$\omega_c \dots\dots\dots$ & 0.1193$\pm$0.0026 & 0.1217$\pm$0.0033 & 0.1112$\pm$0.0055 & 0.1194$\pm$0.0015 \\
			$n_s \dots\dots\dots$ & 0.9653$\pm$0.0081 & 0.939$\pm$0.023 & 0.986$\pm$0.018 &  0.9620$\pm$0.0049 \\
			$\log\left( 10^{10}\,A_s \right)$ & 3.097$\pm$0.036 & 3.08$\pm$0.040 & 3.11$\pm$0.040 &  3.11$\pm$0.034 \\
			$\tau \dots\dots\dots..$ & 0.082$\pm$0.018 & 0.080$\pm$0.019 & 0.078$\pm$0.019 & 0.088$\pm$0.017 \\
			$H_0 \dots\dots\dots$ & 67.7$\pm$1.3 & 66.0$\pm$1.7 & 72.1$\pm$3.1 &
			67.4$\pm$0.71 \\
			$\Delta^2_{\rm rms} \dots\dots.$ & 0.0058$\pm$0.0071 & $-$0.023$\pm$0.020 & 0.040$\pm$0.023 & 0.0009$\pm$0.0050 \\
			\hline
		\end{tabular}
		\caption{Best-fit values and standard deviations for cosmological parameters with the three different Planck data sets (TT, TE and EE polarizations for $\ell>30$), as well as combining them. 
			We have used a prior in $\tau$ in addition to all the data sets to break the degeneracy between $\tau$ and $A_s$.}
		\label{tab:Planck}
	\end{table*}
	
\begin{figure*}[htbp!]
		\includegraphics[width=0.31\linewidth]{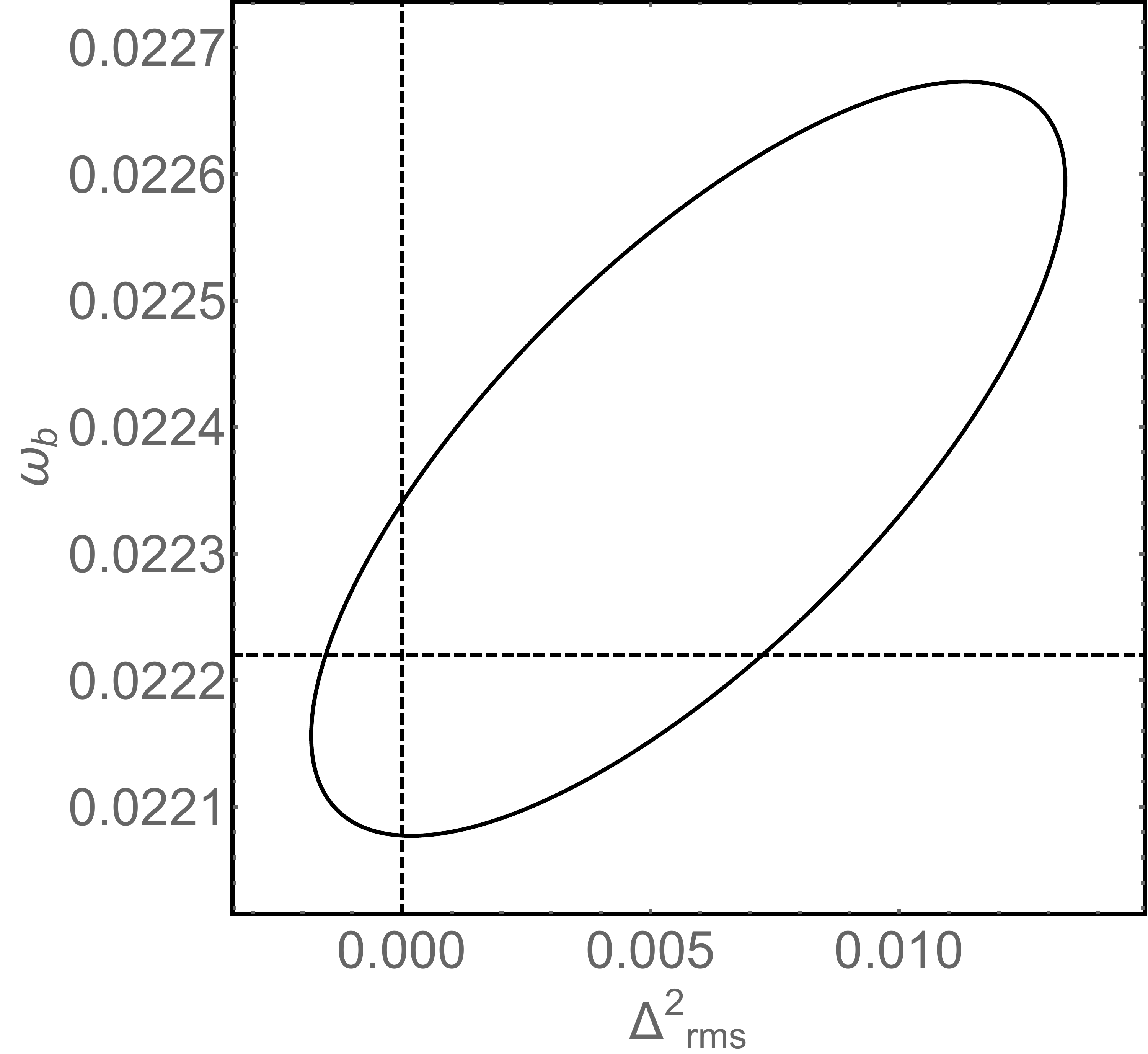}
		\includegraphics[width=0.3\linewidth]{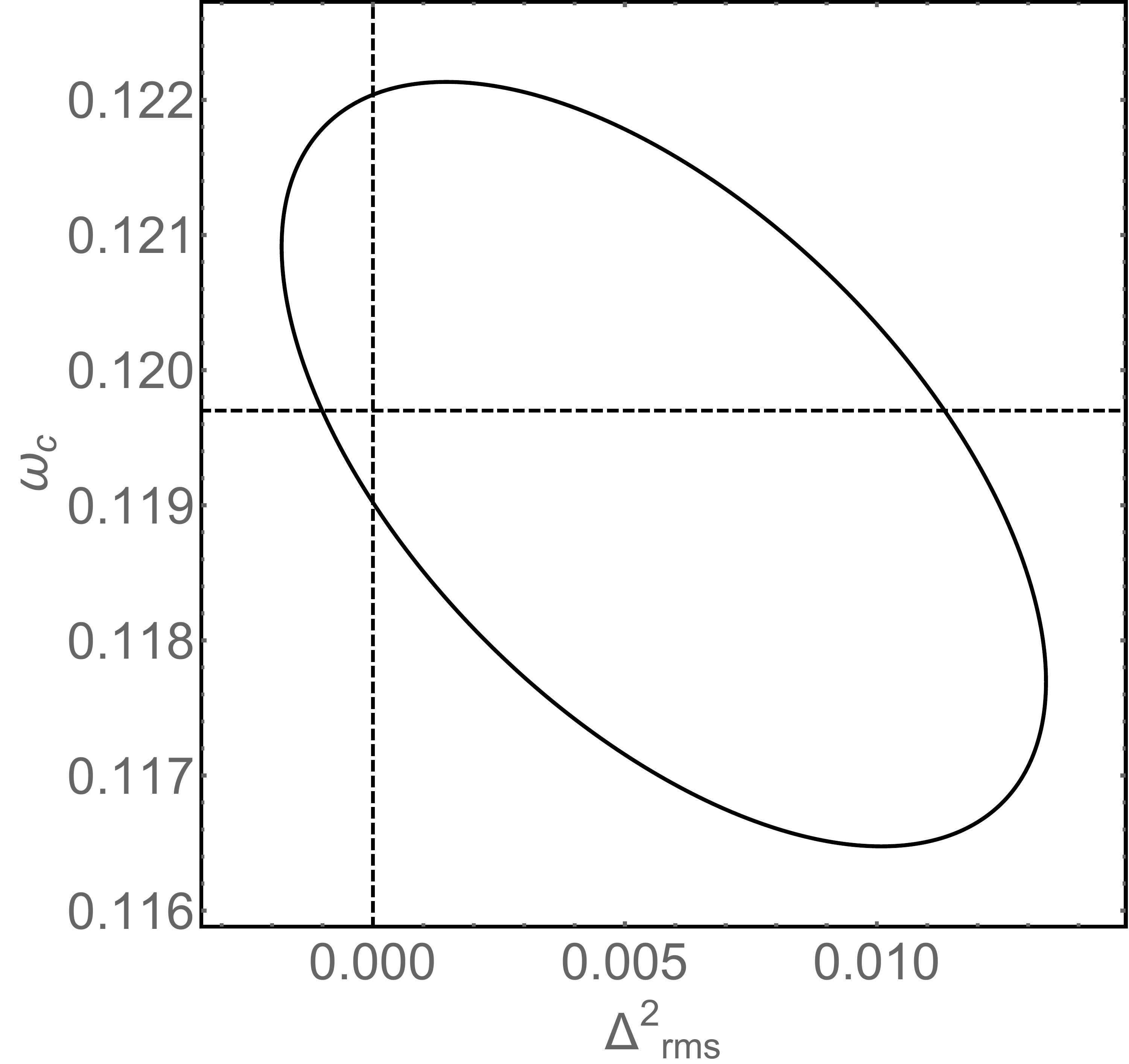}
		\includegraphics[width=0.3\linewidth]{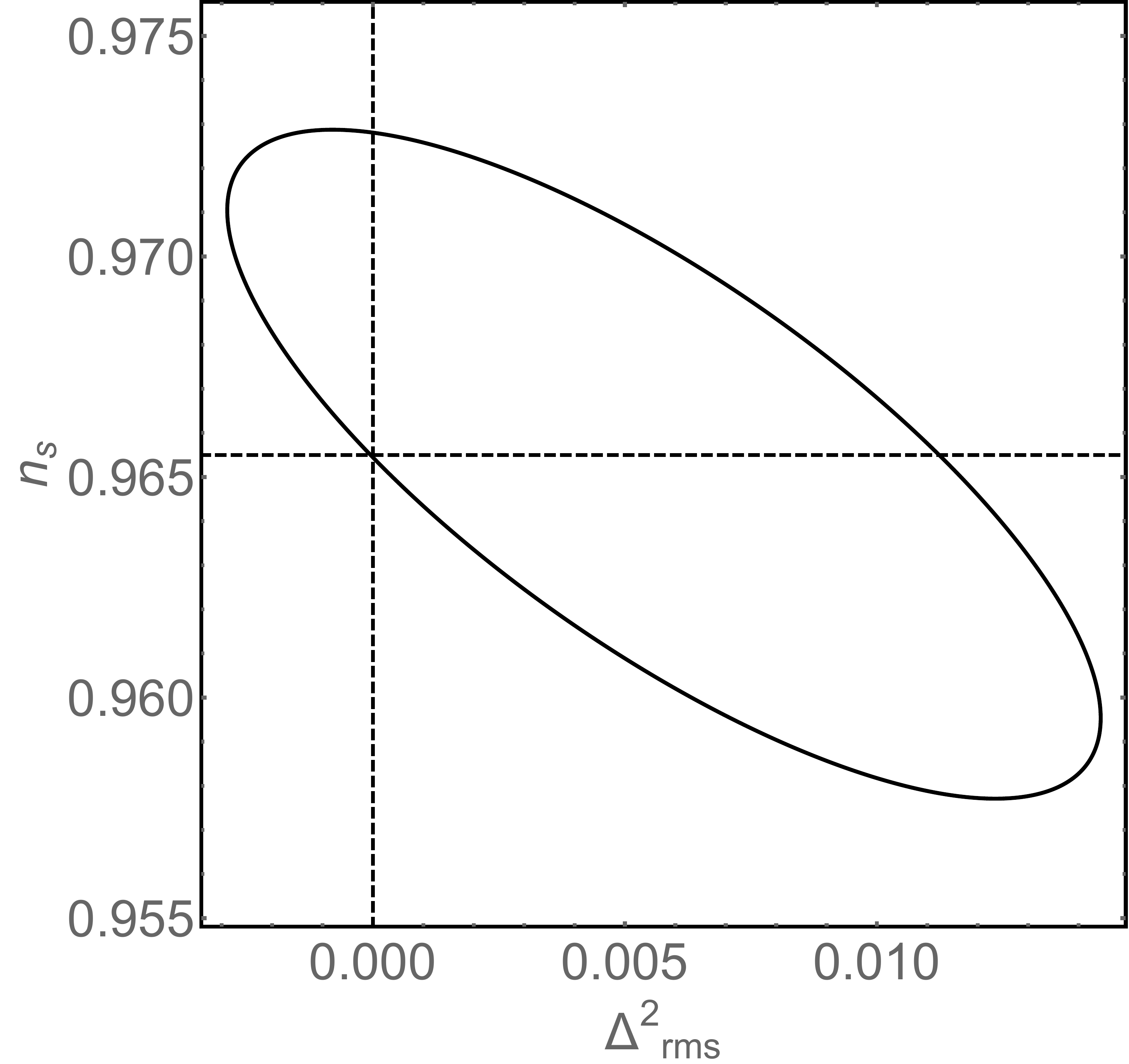}
		\caption{1$\sigma$ (68\%) confidence ellipses for the Planck TT data set. From left to right we show $\omega_b$, $\omega_c$, and $n_s$ vs $\Delta_{\rm rms}^2$. The unperturbed (Planck) best-fit values are shown as dashed lines.}
		\label{fig:ellipse}
	\end{figure*}

\begin{figure*}[htbp!]
		\includegraphics[width=0.295\linewidth]{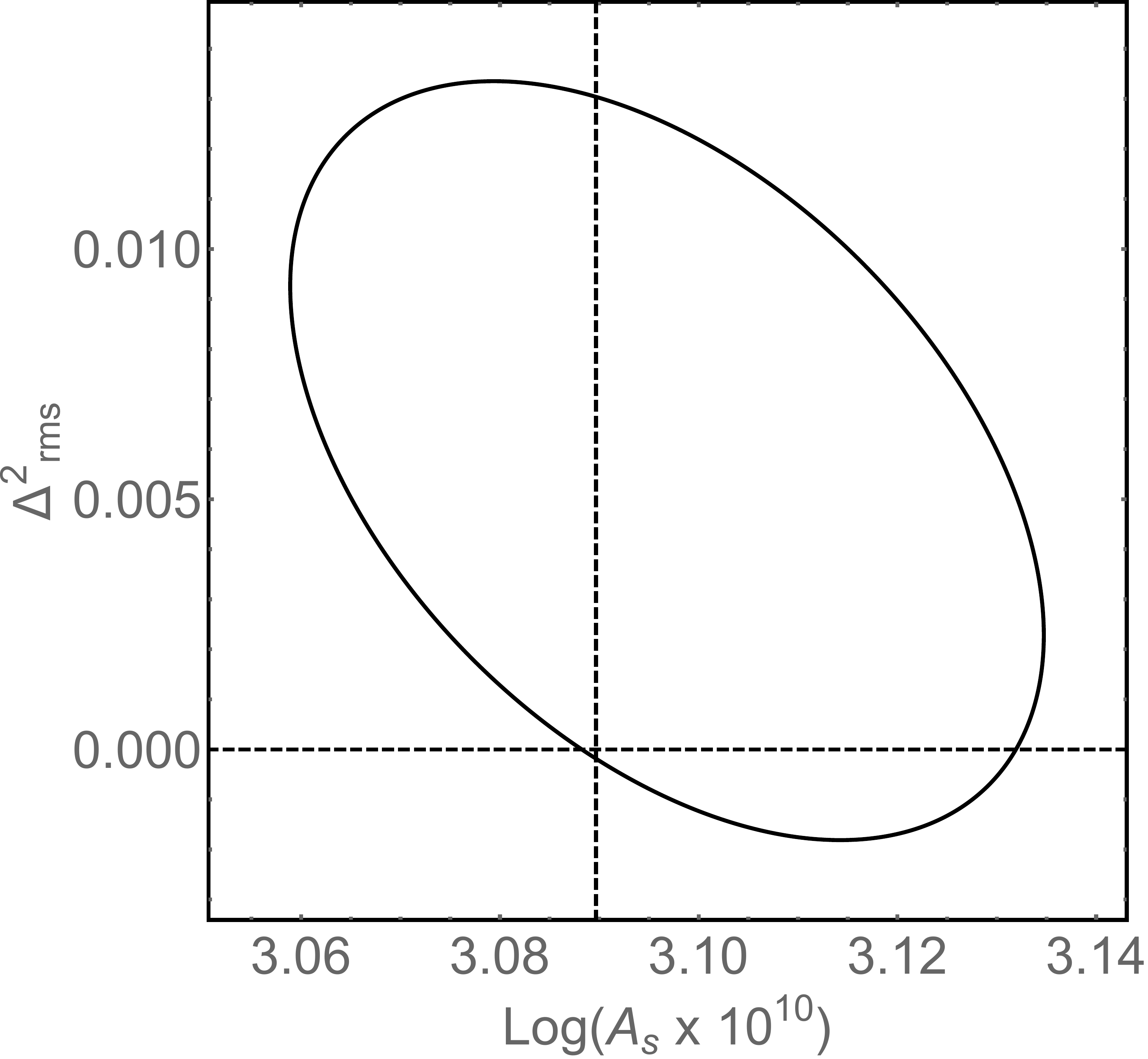}
		\includegraphics[width=0.3\linewidth]{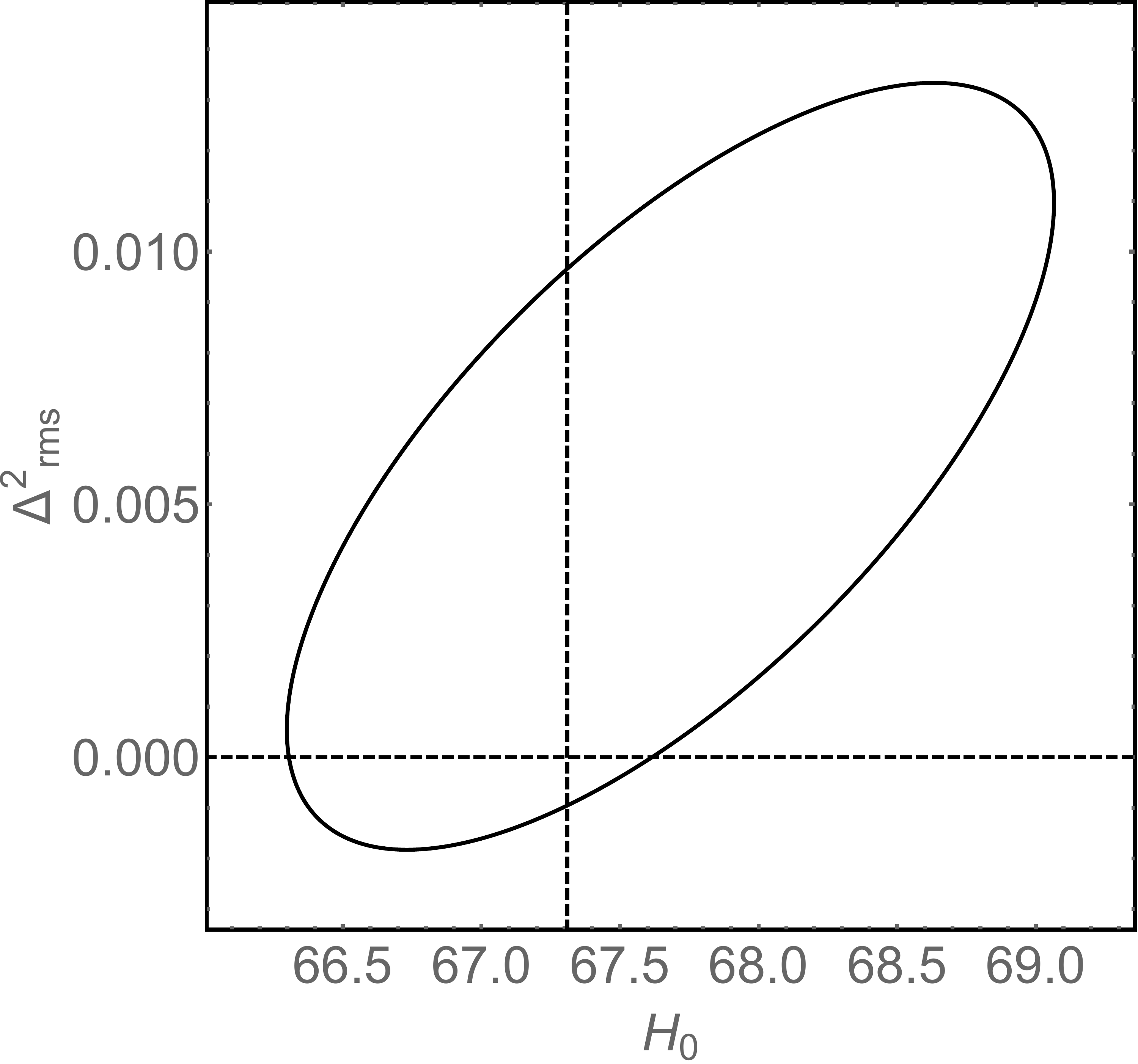}
		\includegraphics[width=0.31\linewidth]{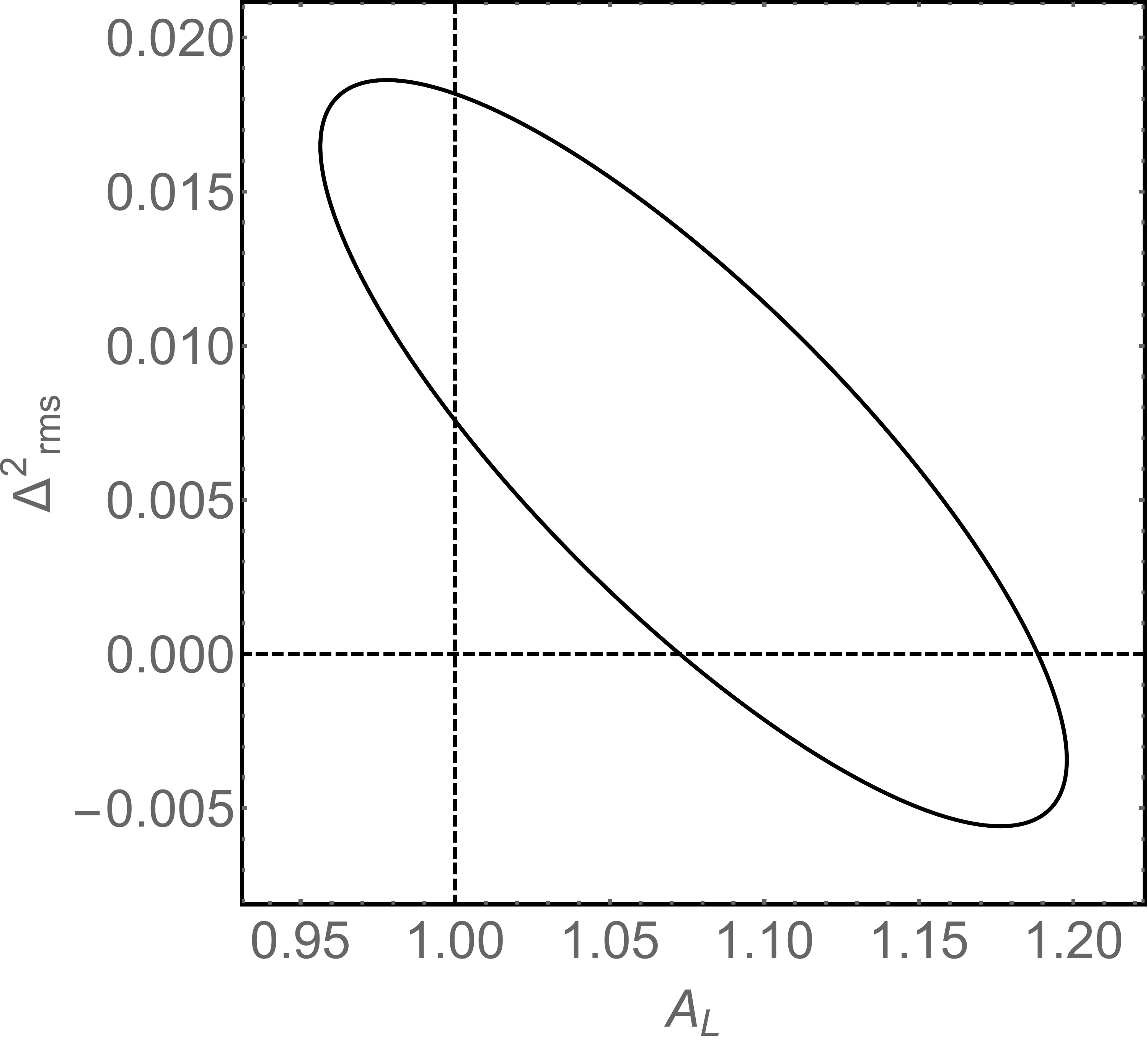}
		\caption{1$\sigma$ (68\%) confidence ellipses for the Planck TT data set. From left to right we show $\Delta_{\rm rms}^2$ vs $A_s$, $H_0$, and $A_L$, using low-$\ell$ polarization data instead of a prior on $\tau$ for the latter. The unperturbed (Planck) best-fit values are shown as dashed lines.}
		\label{fig:ellipse2}
	\end{figure*}

	We show the confidence ellipses for the Planck experiment on Figures~\ref{fig:ellipse} and~\ref{fig:ellipse2}, where it is clear that the CIP contribution to the CMB power spectrum is highly correlated with most of the rest of parameters. The correlation coefficients, defined as  $r_{ij}\equiv (F^{-1})_{ij}/\sqrt{(F^{-1})_{ii} (F^{-1})_{jj}}$, are found to be $r_{\om_b, \Delta^2_{\rm rms}}=0.73$, $r_{\om_c, \Delta^2_{\rm rms}}=-0.57$,  $r_{n_s, \Delta^2_{\rm rms}}=0.76$,  $r_{A_s, \Delta^2_{\rm rms}}=-0.46$, 
	and $r_{H_0, \Delta^2_{\rm rms}}=0.69$.
	
	We do not show the covariance between $\Delta^2_{\rm rms}$ and $\tau$, since the prior applied to $\tau$ renders meaningless the correlation coefficients. Even though this high-$\ell$ analysis shows no strong evidence for the existence of CIPs, they have the potential to resolve the lensing tension mentioned above, when including low-$\ell$ data. We now explore this possibility.

	\subsubsection*{Lensing}
	
	The CMB is lensed by large-scale structure along the line of sight. The main effects of the lensing on the CMB power spectra are to add power at small scales and to smooth the acoustic peaks \cite{astro-ph/0601594,1502.01591}.
	The amount of lensing inferred from CMB TT measurements seems, however, to be higher (by about two standard deviations) than the  predicted value. This difference is parametrized through the lensing amplitude $A_L$ \cite{0803.2309,astro-ph/9803150}, which is fixed to be $A_L=1$ in $\Lambda$CDM, but letting it vary can  better fit the data. An analysis of the  Planck measurements of the TT power spectrum found a best-fit value of $A_L = 1.22 \pm 0.10$ \cite{1502.01589}. 
	
	Adding a new parameter to the likelihood analysis changes the best-fit $A_L$ if the new parameter is correlated with it \cite{DiValentino:2015ola,DiValentino:2015bja}.
	The effects on the CMB of increasing $A_L$ are very similar to adding CIPs, as can be seen from Figures~\ref{fig:gi}, \ref{fig:giTE}, and \ref{fig:giEE}.
	Then, we can compute the offset induced in $A_L$ due a non-zero CIP variance $\Delta_{\rm rms}^2$ as
	\be
	\delta A_L = (F^{-1})_{A_L, \Delta} \, F_{\Delta,\Delta} \, \Delta_{\rm rms}^2.
	\label{eq:biasAL}
	\ee

	In the Planck TT case, the product $(F^{-1})_{A_L, \Delta} \times F_{\Delta,\Delta}$ evaluates to be $\approx -150$. This means that a CIP variance of 
	$\Delta^2_{\rm rms} \approx 10^{-3}$ would induce a bias in the lensing amplitude of $\delta A_L \approx -0.2$, completely eliminating the tension between the $\Lambda$CDM value of $A_{L}=1$ and the observed value. This value of $\Delta^2_{\rm rms}$ is allowed by the current constraints on CIPs, being a factor of $\sim 7$ smaller than our TT-only bound.

	Of course this is only an approximate analysis ignoring the rest of the cosmological parameters. To include all correlations we use a Fisher-matrix analysis as above, adding $\delta A^{X}_{8}\equiv A_L-1$ as an eighth parameter in our analysis. Its associated basis function, $g_8$, is defined as in Eq.~(\ref{eq:gi}).
	
	We fit for the value of $A_L$ from the Planck data, first without CIPs (to show the tension) and then with CIPs. To follow more closely the analysis carried out by Planck \cite{1502.01589}, we will use 
	the low-$\ell$ polarization data instead of setting a prior for $\tau$. These data are available as part of the Planck likelihood package.\footnote{http://wiki.cosmos.esa.int/} 
	
	The results are displayed in Table~\ref{tab:lensing}. We show the fit for the six original $\Lambda$CDM parameters $+A_L$ first, where it is clear that the best-fit lensing amplitude deviates $\sim 2$ standard deviations from the $\Lambda$CDM value of $A_L=1$ for the TT, EE, and the combined data set. 
	
\begin{figure}[htbp!]
			\includegraphics[width=0.9\linewidth]{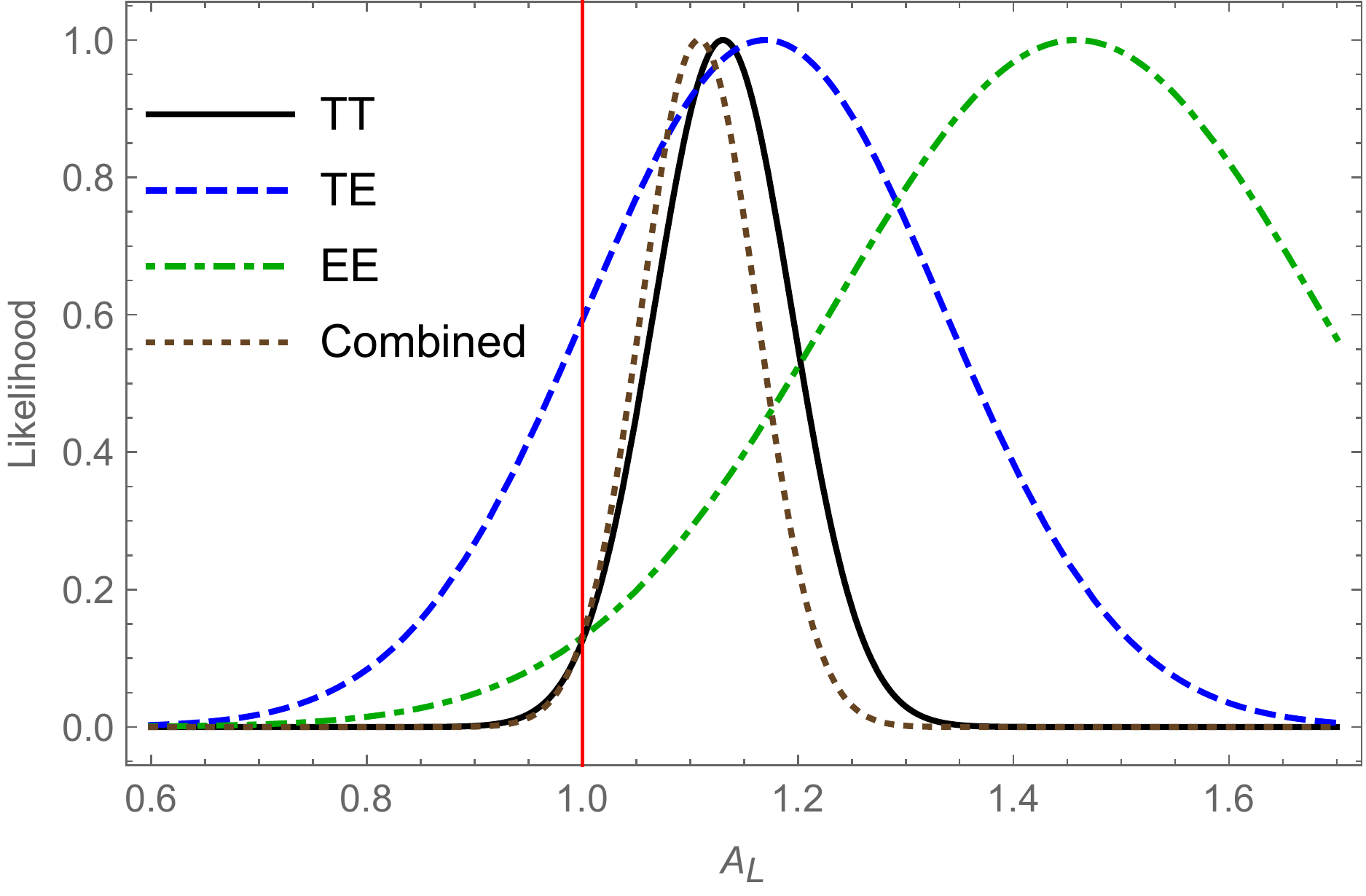}
			\\
			\includegraphics[width=0.9\linewidth]{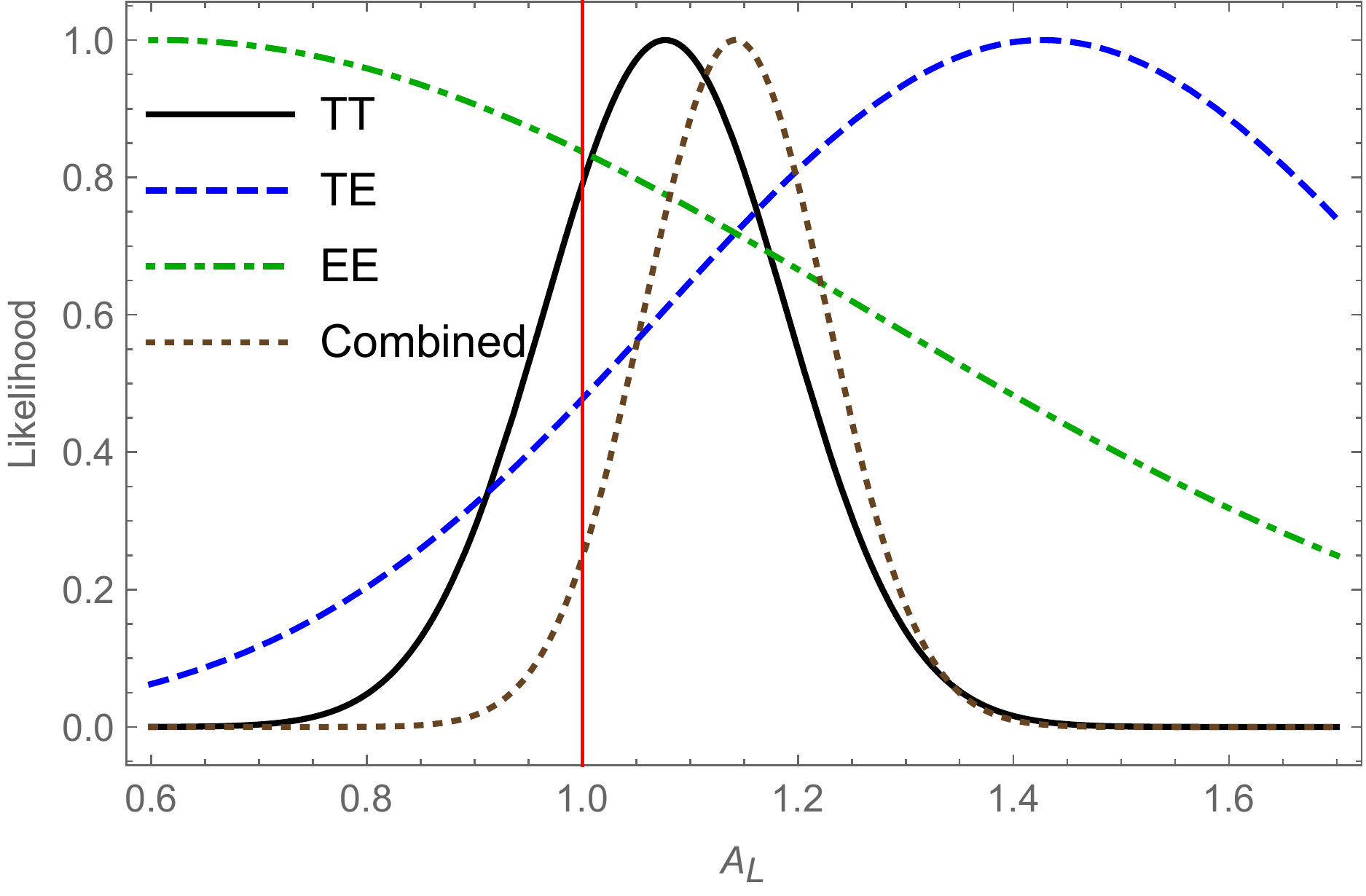}	
			\caption{Normalized likelihoods for the lensing amplitude for all data sets. In the top panel without considering CIPs, and in the lower panel adding them as well.}
			\label{fig:Alens}
		\end{figure}

	In Figure~\ref{fig:Alens} we plot the likelihoods for $A_L$, when marginalizing over the rest of parameters, before and after including CIPs.\footnote{Note that, since we are using a linear Fisher-matrix analysis, these likelihoods are Gaussian by construction.} This Figure shows a significant widening of the likelihood curves, which added to the bias from Eq.~(\ref{eq:biasAL}) is responsible for the decrease in the tension of the fit.
	
	In Table~\ref{tab:lensing} we also show the standard deviations (and new best-fit values) when including the six $\Lambda$CDM parameters + $\Delta^2_{\rm rms} \, +A_L$ (so $N_p=8$). In that case the tension in the TT data set vanishes, due to the correlations between $A_L$ and $\Delta^2_{\rm rms}$.
	
	
	A $\chi^2$ analysis of the TT power spectrum shows a preference for a non-standard lensing amplitude.	
	The change in $\chi^2$  from the standard $\Lambda$CDM model (with $A_L=1$) to an $A_L$-varying model (usually denoted $\Lambda$CDM+$A_L$) is $\Delta \chi^2 = -4.1$, giving rise to a $p$-value of 0.043, which makes it a significantly better fit. 
	
	Adding CIPs to this  $\Lambda$CDM+$A_L$ model changes $\chi^2$ by $\Delta \chi^2 = -0.3$, with a $p$-value of 0.58. This implies that $\Lambda$CDM+$A_L+$CIPs does not fit the TT power spectrum better than $\Lambda$CDM+$A_L$.
	
	Interestingly, CIPs alone can do as well as $A_L$ alone improving the $\chi^2$ statistic. The change in $\chi^2$ from the $\Lambda$CDM model to $\Lambda$CDM+CIPs is $\Delta \chi^2 = -3.9$, with a $p$-value of 0.048 (to be compared with 0.043 when adding a varying $A_L$ to $\Lambda$CDM). Notice, though, that the best-fit CIP variance in that case would be $\Delta^2_{\rm rms} = (12.9 \pm 6.4) \times 10^{-3}$, which is in tension with both the trispectrum bound \cite{1306.4319}, and the galaxy-cluster bound \cite{0907.3919}. 
	
	This shows that adding either a varying $A_L$ or CIPs to a standard $\Lambda$CDM model provides a better fit for the TT Planck power spectrum, by a comparable amount. Adding both, however, is not supported by the data.
	There are, however, a few systematic effects in the analysis that could bias the result. The most important example is that
	our treatment of the low-$\ell$ data is too simplistic. As a result, the uncertainties in $\tau$ and $A_L$ in Table~\ref{tab:lensing} are small when compared to the Planck 2015 result \cite{1502.01589}. This indicates that our Fisher-matrix analysis is too optimistic when inferring the optical depth from the low-$\ell$ polarization data, which could be due to the non-gaussian nature of the low-$\ell$ likelihoods, to the mode coupling, or to the linear approximation breaking down. A full likelihood analysis could show that CIPs absorb more of the lensing tension than indicated in this simple analysis.
	
	Summarizing, we conclude that CIPs are unlikely to solve the lensing tension with current Planck data. 
	Nonetheless, they remain one of the simplest prospective solutions,
	due to their high correlation with the lensing amplitude ($r_{A_L, \Delta^2_{\rm rms}}=-0.82$).
	High-quality low-$\ell$ polarization data will be publicly available in the next few years \cite{1407.2584,1408.4789}, so a reanalysis using the full Planck likelihoods, perhaps also including higher-$\ell$ multipoles from SPTpol \cite{Austermann:2012ga,Crites:2014prc}, will resolve the matter definitively.

	\begin{center}
		\begin{table*}[hbtp!]
			\begin{tabular}{ l  c  c  c  c  }
				\hline
				\hline
				Parameter &  TT & TE & EE & Combined   \\             
				\hline
				$\omega_b \dots\dots\dots$ & 0.02235$\pm$0.00020 & 0.02257$\pm$0.00032 & 0.0245$\pm$0.0013 & 0.02228$\pm$0.00014 \\
				
				$\omega_c \dots\dots\dots$ & 0.1180$\pm$0.0023 & 0.1168$\pm$ 0.0023 & 0.1095$\pm$0.0053 & 0.1185$\pm$0.0014 \\
				
				$n_s \dots\dots\dots$ & 0.9660$\pm$0.0054 & 0.983$\pm$0.015 & 0.992$\pm$0.014 &  0.9641$\pm$0.0037 \\
				
				$\log\left( 10^{10}\,A_s \right)$ & 3.038$\pm$0.031 & 3.065$\pm$ 0.035 & 3.07$\pm$0.034 &  3.038$\pm$0.031 \\
				
				$\tau \dots\dots\dots..$ & 0.056$\pm$0.016 & 0.062$\pm$0.016 & 0.062$\pm$0.016 & 0.055$\pm$0.015 \\
				
				$H_0 \dots\dots\dots$ & 68.1$\pm$1.0 & 68.6$\pm$1.1 & 72.4$\pm$2.9 & 
				67.75$\pm$0.64 \\
				$A_L \dots\dots\dots$ & 1.13$\pm$0.064 & 1.17$\pm$0.17 & 1.46$\pm$0.23 & 
				1.108$\pm$0.054 \\
				\hline
				\hline
				Parameter &  TT & TE & EE & Combined   \\             
				\hline
				$\omega_b \dots\dots\dots$ & 0.02248$\pm$0.00029  & 0.0223$\pm$0.0046 & 0.02593$\pm$0.00017 &  0.02222$\pm$0.00018 \\
				
				$\omega_c \dots\dots\dots$ & 0.1176$\pm$0.0025  & 0.1183$\pm$0.0029 & 0.1102$\pm$ 0.0053 & 0.1187$\pm$0.0015 \\
				
				$n_s \dots\dots\dots$ & 0.9699$\pm$0.0086  & 0.977$\pm$0.017 & 1.004$\pm$0.016 &  0.9622$\pm$0.0052 \\
				
				$\log\left( 10^{10}\,A_s \right)$ & 3.041$\pm$0.031  & 3.05$\pm$0.035 & 3.081$\pm$ 0.036&  3.037$\pm$0.031 \\
				
				$\tau \dots\dots\dots..$ & 0.057$\pm$0.016  & 0.058$\pm$0.016 & 0.065$\pm$0.016 &
				0.054$\pm$0.015 \\
				
				$H_0 \dots\dots\dots$ & 68.5$\pm$1.2 & 67.7$\pm$1.5 & 73.0$\pm$3.0 & 
				67.60$\pm$0.70 \\
				
				$A_L \dots\dots\dots$ & 1.07$\pm$0.11  & 1.43$\pm$0.35 & $-0.39\pm$0.65 & 
				1.142$\pm$0.085 \\
				
				$\Delta^2_{\rm rms} \dots\dots.$ & 0.007$\pm$0.011  & $-$0.028$\pm$0.033 & 0.088$\pm$0.064 & $-$0.0038$\pm$0.0074 \\
				\hline
			\end{tabular}
			\caption{Best-fit values and standard deviations for cosmological parameters with the three different Planck data sets (TT, TE and EE polarizations for $\ell>30$), as well as combining them. 
				In the top part we have fitted for the original six parameters and the lensing amplitude $A_L$. In the bottom part we have also added a CIP amplitude $\Delta_{\rm rms}^2$. Instead of a prior in $\tau$ we have used the low-$\ell$ polarization data ($\ell<30$) from Planck in addition to all the data sets to disentangle $\tau$ and $A_s$.}
			\label{tab:lensing}
		\end{table*}
	\end{center}

	\subsection{Cosmic-variance limit}
	
	We now find the minimum $\Delta_{\rm rms}^2$ observable in the cosmic-variance limited case for different data sets.
	
	We consider an experiment with no instrumental noise $N_\ell$ (i.e. $N_{\ell}=0$), full sky coverage ($\fsky=1$) and range of observation from $\ell=2$ to 2500. In reality the lowest multipoles should be treated with care, due to possible Galactic-foreground subtraction \cite{1507.02704}, which we ignore here. We show the results for the uncertainties of such an experiment in Table~\ref{tab:CVL}. 
	
	The best CVL constraints to $\Delta^2_{\rm rms}$ arise from the polarization power spectra (EE especially) instead of the TT power spectrum, as holds true for the six original parameters \cite{1403.5271}.
	
	The minimum CIP variance observable in the CVL is $ \Delta_{\rm rms}^2 = 9 \times 10^{-4}$, a factor of $\sim$ 5 better than the current trispectrum constraint~\cite{1306.4319}. This result pales in comparison to the sensitivity of a CVL trispectrum experiment, as described in Ref.~\cite{1505.00639}, which would be able to measure $\Delta_{\rm rms}^2 \leq 3\times 10^{-6}$.

	\begin{table*}[hbtp!]
		\begin{tabular}{| l | c | c | c | c | c | c | c |  }
			\hline
			Data &  $\omega_b$ & $\omega_c$ & $n_s$ & $A_s$ & $\tau$ & $H_0$ &  $\Delta^2_{\rm rms}$   \\             
			\hline
			TT &  1.6 $\times 10^{-4}$ & 1.7 $\times 10^{-3}$ & 5.0 $\times 10^{-3}$ & 	8.1 $\times 10^{-11}$ & 0.019 & 0.80 & 4.8 $\times 10^{-3}$ \\
			TE  & 1.1 $\times 10^{-4}$ & 1.0 $\times 10^{-3}$ & 4.7 $\times 10^{-3}$ & 	3.6 $\times 10^{-11}$ & 8.3 $\times 10^{-3}$ & 0.45 & 2.5 $\times 10^{-3}$ \\
			EE & 7.4 $\times 10^{-5}$ & 7.6 $\times 10^{-4}$ & 2.8 $\times 10^{-3}$ & 	9.9 $\times 10^{-12}$ & 2.4 $\times 10^{-3}$ & 0.33 & 1.5 $\times 10^{-3}$ \\
			Combined & 2.8 $\times 10^{-5}$ & 4.6 $\times 10^{-4}$ & 1.8 $\times 10^{-3}$ & 	8.0 $\times 10^{-12}$ & 1.9 $\times 10^{-3}$ & 0.19 &  9.0 $\times 10^{-4}$ \\			
			\hline			
		\end{tabular}
		\caption{Standard deviations forecast for a CVL experiment measuring from $\ell=2$ to $\ell=2500$ and with $\fsky=1$. We consider the six $\Lambda$CDM parameters + $ \Delta_{\rm rms}^2$ being fitted at the same time.}
		\label{tab:CVL}
	\end{table*}

	Here $\tau$ is free, unlike the Planck case, where we included a prior. This leads to higher correlations of the CIP amplitude $\Delta^2_{\rm rms}$ with the optical depth $\tau$, and the scalar amplitude $A_s$.
	We find the correlation coefficients in the CVL case to be $r_{\om_b, \Delta^2_{\rm rms}}=0.30$, $r_{\om_c, \Delta^2_{\rm rms}}=-0.02$,  $r_{n_s, \Delta^2_{\rm rms}}=0.47$,  $r_{A_s, \Delta^2_{\rm rms}}=-0.71$,  $r_{\tau, \Delta^2_{\rm rms}}=-0.66$,
	and $r_{H_0, \Delta^2_{\rm rms}}=0.15$. When including a lensing amplitude, we find $r_{A_L, \Delta^2_{\rm rms}}=-0.27$.

	\section{Conclusions}
	\label{sec:conclusions}

	Compensated isocurvature perturbations leave no imprint on the observable CMB to linear order, so their amplitude can be considerably larger than the $\sim10^{-5}$ amplitude of primordial adiabatic perturbations. Currently the best  constraints arise from analyzing the four-point function of the CMB, from where one can probe the first $L\sim$ 20 multipoles of a CIP power spectrum, corresponding to scales larger than 10 degrees in the sky.
	
	We use a different method to search for CIPs, based on studying the CMB power spectrum that arises to second order in the CIP-perturbation amplitude. We find a simple form for the contribution to the CMB power spectrum, proportional to the CIP variance $\Delta^2_{\rm rms}$, which has the advantage of being easier to analyze than the trispectrum.
	
	The amplitude $\Delta_{\rm rms}^2$ of this new contribution to the power spectrum can be expressed in terms of a sum over all the modes of a scale-invariant CIP power spectrum, although only the first $L\sim 100$ modes are important in CMB studies. This allows us to probe the CIPs down to angular scales of $\sim$ 2 degrees in the sky.
	
	We show that CIPs can alleviate the 2$\sigma$ discrepancy in the lensing amplitude $A_L$, between that inferred from the Planck TT power spectrum and the $\Lambda$CDM expectation ($A_L=1$). Adding CIPs to a standard $\Lambda$CDM model can improve the fit of the TT power spectrum as much as adding a varying $A_L$, making it unnecessary to have $A_L\neq1$.
	The best-fit value for $\Delta^2_{\rm rms}$ in that case, however, would be three standard deviations above the current bounds.
	A full MCMC analysis would precisely determine whether CIPs provide a viable solution to the lensing tension.
	
	We find a 1$\sigma$ constraint on the CIP variance of $\Delta^ 2_{\rm rms}\leq 7.1 \times 10^ {-3}$ using Planck temperature data alone, which improves to $\Delta^ 2_{\rm rms}\leq 5.0 \times 10^ {-3}$ if polarization data are included. 
	This result is of the same order of magnitude as the current trispectrum bound, but this analysis is far simpler and more intuitive, as well as less computationally costly. 
	We also forecast the minimum CIP amplitude observable with a cosmic-variance-limited measurement of the CMB power spectra to be $\Delta^ 2_{\rm rms} \leq 9.0 \times 10^{-4}$. This result is a factor of $\sim 5$ better than the current constraints, promising ever more precise constraints of the uniformity of the primordial baryon fraction.

	\begin{acknowledgments}
		We thank David Spergel, Tristan Smith, Graeme Addison, Marius Millea, Wayne Hu, and Silvia Galli for very useful and instructive discussions, as well as the anonymous referee for helpful comments. JBM, LD, MK, and EDK are funded at Johns Hopkins University by the John Templeton Foundation, the Simons Foundation, NSF grant PHY-1214000, and NASA
		ATP grant NNX15AB18G. DG is funded at the University of Chicago by a National Science Foundation Astronomy and Astrophysics Postdoctoral Fellowship under Award NO. AST-1302856. This work was supported in part by the Kavli Institute for Cosmological Physics at the University of Chicago through grant NSF PHY-1125897 and an endowment from the Kavli Foundation and its founder Fred Kavli. LD is also supported at the Institute for Advanced Study by NASA through Einstein Postdoctoral Fellowship grant number PF5-160135 awarded by the Chandra X-ray Center, which is operated by the Smithsonian Astrophysical Observatory for NASA under contract NAS8-03060.
	\end{acknowledgments}

\end{document}